\documentclass[letterpaper]{article} % DO NOT CHANGE THIS

\usepackage{aaai25}  % DO NOT CHANGE THIS
\usepackage{times}  % DO NOT CHANGE THIS
\usepackage{helvet}  % DO NOT CHANGE THIS
\usepackage{courier}  % DO NOT CHANGE THIS
\usepackage[hyphens]{url}  % DO NOT CHANGE THIS
\usepackage{graphicx} % DO NOT CHANGE THIS
\usepackage{natbib}  % DO NOT CHANGE THIS AND DO NOT ADD ANY OPTIONS TO IT
\usepackage{caption} % DO NOT CHANGE THIS AND DO NOT ADD ANY OPTIONS TO IT
\usepackage{algorithm}
\usepackage{algorithmic}
\usepackage{newfloat}
\usepackage{float}
\usepackage{xcolor}
\usepackage{listings}
\usepackage{booktabs}

\usepackage{newfloat}
\usepackage{listings}
\DeclareCaptionStyle{ruled}{labelfont=normalfont,labelsep=colon,strut=off} % DO NOT CHANGE THIS
\floatstyle{ruled}
\newfloat{listing}{tb}{lst}{}
\floatname{listing}{Listing}
\title{Computational Thinking with Computer Vision: Developing AI Competency in an Introductory Computer Science Course}
\author {
    Tahiya Chowdhury
}
\affiliations{
    Department of Computer Science, Colby College, Waterville, Maine, 04901\\
    tahiya.chowdhury@colby.edu
}

\begin{document}

\maketitle

\begin{abstract}
Developing competency in artificial intelligence is becoming increasingly crucial for computer science (CS) students at all levels of the CS curriculum. However, most previous research focuses on advanced CS courses, as traditional introductory courses provide limited opportunities to develop AI skills and knowledge. This paper introduces an introductory CS course where students learn computational thinking through computer vision, a sub-field of AI, as an application context. The course aims to achieve computational thinking outcomes alongside critical thinking outcomes that expose students to AI approaches and their societal implications. Through experiential activities such as individual projects and reading discussions, our course seeks to balance technical learning and critical thinking goals. Our evaluation, based on pre-and post-course surveys, shows an improved sense of belonging, self-efficacy, and AI ethics awareness among students. The results suggest that an AI-focused context can enhance participation and employability, student-selected projects support self-efficacy, and ethically grounded AI instruction can be effective for interdisciplinary audiences. Students' discussions on reading assignments demonstrated deep engagement with the complex challenges in today's AI landscape. Finally, we share insights on scaling such courses for larger cohorts and improving the learning experience for introductory CS students.
\end{abstract}


\section{Introduction}
The desired traits that enable professional opportunities for computer science (CS) graduates and meet societal needs have been central to shaping the goals of a CS curriculum. As outlined in CS2023, ACM/IEEE-CS/AAAI Computer Science Curricula~\cite{10.1145/3626253.3633405}, these traits define a learning path that prepares CS graduates for holistic development, encompassing technical expertise, ethical awareness, and professional skills necessary for meaningful societal contributions. Ideally, this trajectory should cultivate students' competence by building the knowledge, skills, and behaviors essential for achieving their potential and career aspirations through a well-designed sequence of courses.

However, transforming computer science education to support this vision today remains challenging. First, existing courses must be updated to give students more exposure to emerging tools and real-world contexts. Balancing core computer science knowledge with applied experience can enhance learning and employability. Second, CS is a collaborative field poised to transform many areas of life through its applications. Now more than ever, the CS curriculum must prepare students to collaborate across disciplines, self-learn to solve complex problems, and adapt responsibly to evolving technical needs. This lays the foundation for a future workforce whose competence can make a lasting societal impact. More importantly, offering a holistic, service-oriented view of CS can improve student experiences, link the discipline to other fields, and address diversity issues in the field. Introducing popular topics like AI into introductory courses can boost retention, engagement, and early professional opportunities while providing relevant learning contexts.

This paper presents the design and preliminary outcomes of an introductory CS course aimed at addressing these challenges. Using computer vision as an application context, the course fosters competency in computing and AI. Drawing from existing pedagogical methods, we integrated activities that develop professional skills such as communication, collaboration, ethics, self-directed learning, problem-solving, and persistence, which we believe will better prepare students for professional opportunities across disciplines. We discuss the course results and lessons from the first offering, which will inform future iterations. We believe such a course can enhance introductory CS outcomes, build a versatile AI talent pool, and increase participation in computing.

%* CS 2023 focus on competency, real life projects
%* Students preparation for employability
%* improved student experience with interdisciplinary, broad range of projects
%* professional disposition --- some psychology reference how early behavior help people --- first year students need there network
%* Diversity in CS, service to other disciplines

\section{Background and Related Work}

Contexts can leave lasting impressions on students and have proven effective for retention and broadening participation in CS~\cite{10.1145/2493394.2493397}. Previous studies have explored context-based learning strategies, integrating computer science into fields like journalism~\cite{10.1145/2839509.2844636}, music~\cite{10.1145/3017680.3017767}, and biology~\cite{10.1145/3408877.3432469}. A key challenge in designing such interdisciplinary courses is ensuring that students gain computing skills comparable to traditional CS courses without needing to major in computer science. Contexts have also been used in introductory programming courses, integrating engaging CS-related topics like robotics, games, and science into CS1 (or equivalent) courses~\cite{10.1145/2133797.2133801, 10.5555/2184451.2184458, 10.1145/3017680.3017757}, creating personalized contexts for motivation. Our course draws from this approach, aiming to enhance students' sense of belonging in computing and AI.
%Students learn computing concepts using Computer Vision as a context that they can apply to identify AI techniques for developing a computing solution to a real-world problem.

A core aspect of context-based learning is offering hands-on experience through methods like experiential and service-based learning. ~\cite{vandegrift2007encouraging} explored open-ended programming assignments in introductory courses, giving students control over their learning, and reported that students made extra efforts when working on projects related to their personal interests. Similarly, ~\cite{yarosh} found that students engaged more deeply with additional problem sets when they found them enjoyable. For non-CS majors, ~\cite{10.5555/1791129.1791141} implemented personal projects in a MATLAB course, allowing students to apply course concepts to a scientific discipline of their choice. Following this approach, we incorporated a final project where each student tackled a unique problem, applying computing concepts, AI tools, and their personal interests, giving them more agency and improving self-efficacy.

As computing and AI increasingly shape how we live and work, understanding their societal, ethical, and professional implications is crucial. ~\cite{10.1145/3626253.3635557} surveyed educators, highlighting the need for a multidisciplinary approach to improving ethics awareness. ~\cite{10.1145/3341164} reviewed syllabi and proposed course modules to embed ethics in existing assignments. Studies like~\cite{10.1145/3545945.3569802, 10.1145/3478431.3499407, 10.1145/3545945.3569881} reported positive outcomes from embedding ethics modules in intermediate-level courses. ~\cite{10.1145/3626252.3630792} incorporated reading, discussion, and in-class activities for students to reflect on the societal impacts of computing tools. However, teaching AI and computing ethics in introductory courses remains underexplored. Our course integrates reading assignments, discussions, case studies, and debates to help students grasp the ethical implications of widely used AI systems. To assess outcomes, we administered pre- and post-course surveys, tracking changes in students' sense of belonging, self-efficacy, and ethics awareness, building on prior research~\cite{belonging, Lent1986SelfefficacyIT}.

%* How other courses incorporated context based learning
%* CS1 courses introduces students to programming or interest CS, they bring AI interest
%* Exposing them to the perils of chatGPT can easily show them why relying on AI for assignments are detrimental, cannot assume they know
% can one cource for all cater to nonn-major an dCS majors

\section{Methodology}

\subsection{Course Context}

We developed this course to teach computational thinking as an entry point to computer science for first-year undergraduates. Our choice of computer vision, a subfield of Artificial Intelligence, as the application context is based on several factors. First, introductory CS courses are seeing high enrollment, driven by student interest in both CS majors and general computing. Given the diverse backgrounds and career goals, learning objectives in such courses vary, so we designed this course as an advanced form of computational thinking. Here, students can evolve their understanding of computational concepts into practical skills and apply them to problem-solving. Additionally, there is increasing discussion about AI's impact on learning, creativity, and society. By using computer vision as a focus, we aim to make computing knowledge more relevant, helping students connect to how current AI tools are developed and their societal influence. Third, AI is rapidly changing the future of work, reshaping how we live, work, and interact. Learning to use and work with AI tools will prepare students to become informed users of these technologies, enhancing their employability.

\subsection{Course Description}
This course was first offered in the Fall of 2023 at a small undergraduate-focused institution (Colby College, Waterville, Maine). The course fulfills Computer science major requirements (equivalent to CS1). Other CS1-equivalent courses were also offered, but students could not receive credit for both. Enrollment required passing a placement test on basic programming concepts or being identified as 'advanced' by faculty teaching other introductory computing courses. The 100-level course, primarily for first-year students, is a prerequisite for CS2-equivalent courses at our institution. Below is the course description from the institutional course catalog:

\begin{table*}[t!]
  %\vspace{-10pt}
  \centering
  \begin{tabular}{c|c|cccc|c|ccc}
    \toprule
    \textbf{Semester} & \textbf{Female} & \textbf{White}  & \textbf{Asian} & \textbf{Black} & \textbf{Hispanic/Latinx} &\textbf{Non-CS} & \textbf{1st yr.} & \textbf{2nd yr.} & \textbf{3rd yr.}\\
    \midrule
    Fall & 27\% & 40\% & 18\% & 13\% & 8\% & 53\% & 67\% & 10\% & 8\%\\
    %NLP & 38\% & 41\%  & 33\% & 2\%  &  8\% & 64\% & 93\% & 5\% & 1\% \\
    \bottomrule
  \end{tabular}
  \caption{Percentage statistics of survey respondents' demography in the course.}
  \label{tab:demography-summary_fall}
%\vspace{-16pt}
\end{table*}

%CV project
\begin{table*}[t!]
\centering
  %\vspace{-10pt}
  \begin{tabular}{llll}
    \toprule
    No. & Week & CV Project & Tools\\
    \midrule
    1 &  1 & Getting started with notebooks; Image Processing & OpenCV\\
    2 & 2-4 & Image manipulation and augmentation & SciPy\\
    3 & 5 & Final project proposal & \\
    4 & 6-7 & Image clustering with pre-trained CV model & Keras, HuggingFace\\
    5 & 8-9 & Classification with image and video data & MediaPipe, YOLO\\
    6 & 10-13 & Designing an AI application (detection, segmentation, tracking) &  Personal Project \\
    \bottomrule
  \end{tabular}
  \caption{Project schedule for the course.}
  \label{tab:projectCV}
%\vspace{-6pt}
\end{table*}

\textit{An introduction to computational thinking: how we can describe and solve problems using a computer. Using Python programming language, students will learn how to write algorithms, manipulate information, and design programs. They will learn about abstraction, how to divide and organize a process into appropriate components, how to describe processes in a computer language, and how to analyze and understand the behavior of their programs. The projects will focus on manipulating image data using computer vision. This course enables CS and AI-related student learning outcomes and requires prior programming experience.}

\subsection{Learning Outcomes}
This course, which counts toward the CS major, is designed to achieve learning outcomes related to both CS and AI. These outcomes, outlined in the syllabus, shaped the course activities.

\textbf{Computational Thinking.} The course aims to teach students how to approach problems, design solutions using computational concepts, and implement them through programming. In-class programming exercises, weekly quizzes, and homework were designed to meet these computational thinking (CT) goals. By the end of the course, students should be able to understand abstraction, convert problem statements into procedural programming solutions, and effectively communicate their results. Topics included programming fundamentals such as data types, variables, functions, control flow, and object-oriented programming concepts like classes, inheritance, and memory models. These topics were covered in lectures and assessed through weekly homework, quizzes, and a final exam.

\textbf{Critical Thinking.} Recognizing the varied experience levels of students with computational processes and AI, we used a mix of activities, including reading assignments, in-class discussions, group work, and open-ended projects, to build AI competence. By the end of the course, students should be able to critically analyze AI/ML papers and media discussions, participate in group discussions, and understand the legal, cultural, and ethical implications of AI/ML systems on individuals, communities, and society.

\subsection{Course Structure} This course was offered during the Fall session (September-December term) as a 4-credit, 14-week course. Enrollment was capped at 25, and 12 students registered and completed the course. The class met twice a week (Tuesdays and Thursdays) for 75-minute sessions, totaling 150 minutes of class time weekly. The instructor, an early-career CS faculty member with 5+ years of college-level programming teaching experience, was assisted by one teaching staff for grading projects and homework.

Each Tuesday began with a discussion of the previous week’s reading assignments and homework (15–20 minutes), followed by a lecture on programming concepts in Python (25–30 minutes), and concluded with individual or group programming activities. On Thursdays, the class began with a quiz (10–15 minutes) via Moodle, followed by a lecture (25–30 minutes), and finished with open project work time.

\begin{table*}[thp]
  \centering
  \begin{tabular}{|p{.25in}|p{5.0in}| p{0.5in} p{0.5in}|}
    \toprule
      \textbf{Item} & \textbf{Question} & \textbf{Change}  & $p$-value\\
    \midrule
     \hline
     & \textbf{Sense of Belonging} & & ~~~\\
     %\hline
    1 & I feel like I belong in AI. & $+0.6969$ & $0.3380$ \\
    2 & I see myself as an AI tool user/builder. & $+0.2727$ & $0.4375$\\
    3 & I feel like an outsider in AI. (reverse coded) & $\textbf{+0.3939}$ & $<0.05$\\
     \hline
    & \textbf{Self-efficacy} & & ~~~\\
     %\hline
    7 & I am confident that I can choose and use AI tools for CV. & ${+1.1667}$ & $0.140$\\
    8 & I am confident that I understand the procedures and main steps of building an AI tool. & $\textbf{+1.4242}$ & $<0.05$ \\
    9 & I am confident that I can understand and interpret the results/outputs from an AI model. & $\textbf{+1.5909}$ & $<0.05$ \\
    10 & I am confident that my skills in AI will enable me to work with a faculty member on CV tasks. & ${+0.3181}$ & $0.0931$\\
     \hline
    & \textbf{AI Ethics Awareness} & &~~~\\
     %\hline
    11 &  I am comfortable with using AI-driven products in daily life, like smart search, voice assistants, ride sharing apps and social media.& $+0.3181$ & 0.5785
 \\
    12 & I understand and am comfortable with how AI-driven products collect and use my data.& ${+1.1060}$ & $0.2910$\\
    13 & I need to know more about how AI-driven products use my data to become comfortable using them. & $+0.4696$ & $0.2462$ \\
    \bottomrule
  \end{tabular}
  \caption{Survey items are grouped under the three measures used to evaluate the course. Changes in mean response between the initial and final surveys are reported along with the significance of the statistical test. Statistically significant changes are presented in bold letters.}
  \label{tab:survey_fall}
%\vspace{-10pt}
\end{table*}

\subsection{Reading Activities}

To have adequate exposure to the history of computing and AI, our students experienced reading assignments as their first learning activity for the week. We found two benefits from incorporating readings that are excerpts from AI/ML literature, media articles, and videos. First, with the growing concern over Generative AI's~\cite{saetra2023generative} potential impact on computing skills and jobs, we wanted our students to form their own opinions through the readings about the policies, usefulness, and harm of AI misuse. Second, many students bring enthusiasm about learning about AI and misunderstand the widely publicized large language models like ChatGPT~\cite{chatgpt} or Gemini~\cite{team2023gemini} as the only form of AI, unaware of the longstanding AI research and products deployed in our everyday life. Our weekly readings were designed to expose students to the full lifecycle of AI systems, helping them understand when and how to use them responsibly.

Building on prior research on reading assignment engagement via interactive platforms~\cite{10.1145/3017680.3017732}, we used Perusall~\cite{perusall}, an online social annotation tool, to boost student engagement and create a collaborative reading experience. All readings were posted at the start of the semester, allowing students to work at their own pace. We also incentivized participation by awarding points for comments or replies on the readings, which counted toward class participation.

The weekly reading assignments were designed to align with lecture topics. Early assignments involved short videos on general topics (e.g., what is AI, AI vs. data science), followed by a mix of interactive tools, tutorials, journalistic articles, and research papers on AI methods and applications. These readings also addressed concerns about ML datasets and algorithm limitations and critiqued future AI directions. Selections were drawn from top-tier news outlets (e.g., The New York Times, The Guardian) and research forums focused on AI ethics. To balance workloads, we limited readings to two per week, being mindful that students had other courses to manage during the fall semester. \footnote{The full reading list is publicly available upon request.}

The CV topics in this course are not a substitute for a course aimed at advanced-level computer vision students. Instead, we use computer vision as an application context in an introductory class to teach computational thinking, balancing learning outcomes between CS and AI. Given the breadth of CV, we focus on a few topics suitable for beginners through projects and expose students to advanced techniques via readings. We also balance ethical and technical AI content. This approach may raise concerns about insufficient depth in foundational AI topics, such as statistics and linear algebra, which are often prerequisites for advanced AI courses. We prioritize computational thinking as the primary outcome and AI learning as the secondary to manage the breadth-vs.-depth trade-off. While this limits AI topic depth, it enables hands-on AI learning in a 100-level class without prerequisites, aligning with the course’s core objective.

%We used a selected subset of the readings from the winter courses which needed to be completed on a weekly basis. The rationale behind this is students were engaging with other courses in parallel during the fall semester, making it difficult for students to complete a large volume of readings. For both fall courses, students completed readings on Perusall and interacted with each other through comments and replies. Due to limited class time compared to the Fall courses, we did not include follow-up discussions on the readings as an in-class activity.

\textbf{Discussion.} Before each class, the instructor reviewed all comments and selected those that merited in-person discussion. Occasionally, we engaged students in group activities like role-playing (e.g., investigating the reliability of face detection technology in medical facilities) or debating (between ML ethics and engineering teams about addressing data bias). In groups of four, students discussed the topics, with one member designated to take notes. After 10–15 minutes of group discussion, each group presented their key points to the class. The entire discussion session was kept within a 15–20 minute timeframe.

\subsection{Lectures and Programming Activities} The lecture component used self-explanatory Colaboratory notebooks, where programming concepts were introduced with examples. Each notebook included 2-3 programming exercises, allowing students to apply the concepts to solve problems. The exercises focused on basic programming tasks, with students writing programs in Python. Students began working on the exercises during the open work session in class, completed them outside of class, and submitted their work via Moodle by the end of the week as homework.
%During open work hour session of the class, the students worked on these programming problems which were designed to focus on the topic covered that day. 
%For both lecture classes in the week, students had 2-3 practice programming problems to help teach applying the concepts and syntactic knowledge for a relevant context.

\subsection{Projects} Bi-weekly projects focused on applying CS concepts in the context of computer vision. Each student completed these projects individually, using a starter code provided by the instructor. The projects were cumulative, allowing students to reuse code from previous assignments (e.g., project two built on code from project one). Completed projects were submitted on GitHub. Table \ref{tab:projectCV} shows the outlines of the topics of each project. Students used several open-sourced CV libraries (Keras, OpenCV, Mediapipe) to complete the projects.

 \begin{figure*}[!htp]
\centering
    \includegraphics[width=.55\textwidth]{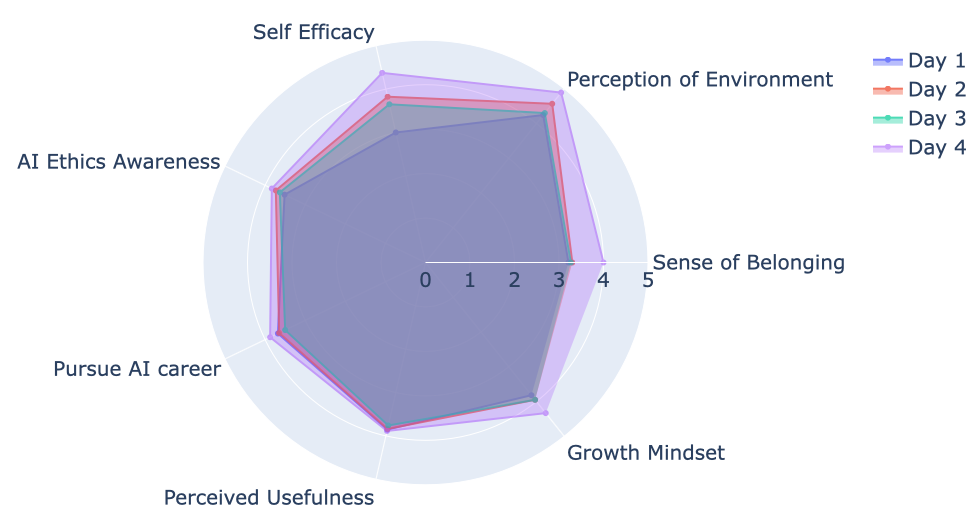}
    \hspace{0.2pt}
    \includegraphics[width=.40\textwidth]{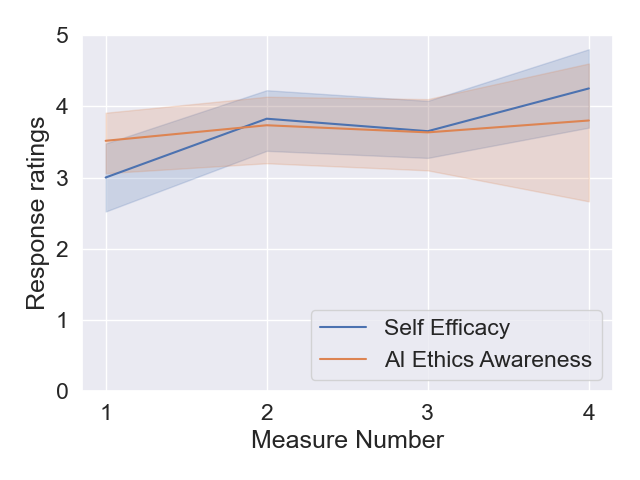}
    \caption{(Left): Radar plot showing aggregate student responses from the four repeated measures over the course. We observed an increase in AI ethics awareness, sense of belonging, and self efficacy.}
    %\Description[Radar plot showing aggregate student responses for the four repeated measures over the Fall semester for the CV.]{We observe an increase in AI Ethics Awareness, Perception of the environment, Sense of belonging and Self Efficacy, and decrease in Growth Mindset.}
\label{fig:radar_plots_Fall}
%\vspace{-8pt}
\end{figure*}

\textbf{Final Project.} A key feature of this course is the Final Project, allowing students to apply their course knowledge to a project of personal significance. We called it a `Personal Project' to encourage students to draw inspiration from their academic or non-academic interests. The workload was divided into three deliverables: 1) the project proposal, 2) a work-in-progress report, and 3) the final milestone. The project proposal, due in Week 6, ensured that the project was feasible within the timeline, involved Python programming and images, and held personal relevance. Students outlined the goal, problem statement, approach, connection to their interests, and references. Feedback on the proposal was used to revise it over two weeks. By Week 11, students submitted their work-in-progress, including a datasheet, model card, references, and a source code notebook with documentation for a non-technical audience. In Week 13, students gave a 5-minute presentation to showcase their project’s features. The final submission included a pre-recorded demonstration video, a written report, and a well-documented source code repository.

%In the final week, students needed to submit a 5-minute demonstration video, a project report, source code, and a final presentation (5 minutes) to share the final version with the class.  

\subsection{Assessment}
The final grade for both courses is composed of: Projects (45\%), Quizzes (15\%), Homework (10\%), Class Participation (10\%), and the Final Exam (20\%). The final exam was administered online via Moodle (the course management site for this course) which included a blend of programming and conceptual questions similar to those on homework and quizzes. For the final project demonstration, we included peer assessment to offer iterative feedback. Each student rated every project in four categories: Creativity, Presentation and Engagement, Technical Work, and Application.

% syllabus
% prior knowledge
% lectures: part 1, part 2
% Homeworks
% quizzes
% Exams
% Labs: Projects
% reading assignments
% Personal projects: image, programming, personal, allow LLMs
% important to know LLMs are not the only AI
% Notable computer scientist video

\section{Data}\label{data}

\subsection{Data Collection}\label{data_collection}

To assess the course's impact on student learning, we administered a pre-survey on Day 1 to establish a baseline. To avoid respondent fatigue~\cite{o2017factors}, we repeated the surveys on the last day of each month, collecting responses four times during the course (with the final survey on Dec 7). Both pre- and post-surveys contained identical items. The survey items were adapted from prior work~\cite{10.1145/3328778.3366805} on sense of belonging and self-efficacy, and AI ethics measures were designed by the authors for the purpose of the study. Each question
Students were informed about the research purpose and given the option to remain anonymous. Demographic data were collected, but no personally identifiable information (e.g., names or emails) was gathered. Surveys were conducted via Google Forms as an in-class activity to incentivize participation when the instructor was present in the class. A summary of the demographic information about the respondents is provided in Table ~\ref{tab:demography-summary_fall}.

\textbf{Survey Measures.} Students were asked about their sense of belonging, self-efficacy in AI, and AI ethics awareness, among other items. Survey items and statistical results comparing the initial (Day 1) and final surveys (Day 4) are detailed in Table~\ref{tab:survey_fall}. Responses were rated on a scale from 1 to 5 to reflect the student's agreement with the statement, where 1) Strongly disagree, 2) Somewhat disagree, 3) Neither disagree nor agree, 4) Somewhat agree, 5) Strongly agree. Note that we reverse-code item 3 for better interpretation of its mean ratings with other items for statistical analysis.

%The survey was accompanied by the following statement:
%\textit{
%\begin{quote}
    %Please rate your level of agreement or disagreement with each of the following statements. We ask these questions in order to assess the efficacy of this course. Your responses will not be linked to your student ID. Your responses will not be used in grading.
%\end{quote}}
% surveys
% quotes

\textbf{Qualitative data.} %In addition to the survey responses, we collect qualitative data on students' thoughts and perspectives. 
To understand how well the students understood the reading materials, we asked students to post at least 2-3 comments on each reading assignment on Perusall. At the end of the semester, we downloaded all comments for the courses into a single database that we use as our source for qualitative data on critical thinking and communication.

\subsection{Data Analysis}

%We analyzed survey responses from each day to generate descriptive statistics (mean, standard deviation) and performed statistical tests to determine the significance of the change observed. 
To compare pre- and post-survey responses, we selected the Wilcoxon rank-sum test over the t-test due to its suitability for ordinal data, small sample sizes, and non-parametric nature. For more interpretive results, we reverse-coded survey item 3 to align with the scale described in section~\ref{data_collection}. For qualitative analysis, we employed inductive coding to conduct a thematic analysis of student comments on reading assignments.

%Omit, if we do BERT-based analysis
 %The first author analyzed the comments using multiple rounds to identify themes, and categorize and summarize emerging patterns to understand student perspectives of the topics.

\textbf{Hypothesis 1}: Students' sense of belonging, self-efficacy, and AI ethics awareness will improve after taking the course. This implies that the mean response on the final survey will be higher than on the initial survey.
%This means, the difference between the mean response of the last survey and the first survey will be positive, and the higher the response is, the better.

%may need to revisit this
\textbf{Hypothesis 2}: Student will demonstrate a critical understanding of various power dynamics in the ML annotation process and responsible AI practices. %This means students should be able to engage in discussion and understand the legal, cultural, and ethical impacts of the AI systems on individuals, groups, and society at large.
%a critique of existing practices in AI and generate thought-provoking questions from a socio-technical perspective.

\section{Results}

%surveys
\subsection{Improved Outcome in Student Perceptions}
Table~\ref{tab:survey_fall} shows the mean changes in student ratings between the pre- and post-surveys. All 13 items demonstrated positive changes, with two self-efficacy items showing statistically significant improvements. Sense of belonging also saw significant positive change. However, AI ethics awareness did not significantly improve, contrary to expectations. This may be due to students' already high levels of AI ethics awareness in the initial survey (see Figure~\ref{fig:radar_plots_Fall}(right), which remained stable throughout the semester.

% quotes from CV166 courses
\subsection{Student Discussions Demonstrate Ethical Awareness}

Next, we analyzed student comments on Perusall in response to the reading assignments. We hypothesized that engaging with these readings would give students a broader understanding of what it takes to develop an AI system.
We highlight several key themes from the student comments here. Regarding ethical conflicts in data collection and annotation discussed in~\cite{nytimes, JMLR:v18:17-234}, one student noted:
\begin{quote}
    \textit{This isn't a conflict I hear brought up very often in discussions of AI. Most of the time people feel threatened by AI and are worried about their employment...While it would be more efficient and accurate for someone who already knows medical skills to quickly identify photos, they might not have the time to do this additional work. So someone who doesn't know anything or very little about the field must do the work. This means they need to be trained and there is more room for mistakes than for a trained professional.}
\end{quote}

Another student holistically summarized the invisible human labor needed to make modern AI work:
\begin{quote}
\textit{My first impression when I read about this data labeling industry is that it's unethical, but it's probably a net good for society. It's created jobs for people from low-income homes in developing countries and it's helping save lives through early detection of symptoms in patients which is good.
...I think the data labeling industry needs to be regulated but not abolished. Companies need to be forced to be as transparent as possible to their customers about what data they collect, and where it ends up. Workers in this industry should also be provided with certain minimum benefits and be paid the full worth of their work by law.}
\end{quote}

Regarding the important role data plays in AI and how its out-of-context use gives rise to biased systems, discussed in~\cite{gebru2021datasheets}, a student reflects:

\begin{quote}
    \textit{The first thought that came to [my] mind after reading this [article] is that %it's possible that this could be caused not by people in the machine learning community simply neglecting to provide information about their datasets but by the fact that 
    machine learning is not a standardized, or straight forward process and is sometimes more of an art than it is a science. It's possible that ML engineers often don't know beforehand what exactly a data set might be useful for at the time it's being created, but then in the future someone comes up with a bright idea for what the data can be used to predict, and it's at that point that the bias in the data and the way it was collected reveals itself. [...] I think documenting the process used for creating the data set would still help significantly.}
\end{quote}

Along with a similar discussion on model documentation for increasing transparency about AI models in~\cite{mitchell2019model}, 

\begin{quote}
    \textit{...This is important because there could be an infinite amount of ways to group intersectional identities, so pointing out intended users with a more specific grouping system will be more accurate [to be useful].}
\end{quote}
Apart from the readings on critical analysis of current practices in AI development, students were also able to learn about classic machine learning concepts used to build state-of-the-art computer vision models, including neural networks, convolutional networks, detection, etc. A student reflected on these core concepts based on their prior ideas about them:

\begin{quote}
    \textit{I thought a neural network was going to be a bit more like a web. This looks more like a function where you put something in and something gets out.}

\end{quote}

%Strikingly, some students put forth questions that are also garnering attention and concerns in the AI research and development community, such as about computing bottleneck:
%\begin{quote}
  %\textit{...It seems that computing is a big part of computer vision as they come with new algorithms. Are we reaching the limit of making AI processes more efficient?}
%\end{quote}

Finally, a student reflected on the energy efficiency and the environmental impact of large AI models:

\begin{quote}
\textit{The more I read about the amount of energy that is required for these models to
be trained, I seriously believe that energy consumption data should be attached to
every ML model. This needs to be a must.}

\end{quote}

% surveys

% response in course review
\subsection{Student Feedback}
%We also looked at the student-provided feedback in the anonymous course evaluation. 
In the course evaluation, all students except one (who responded ``Agree") indicated ``Strongly Agree" that the course was intellectually challenging. Every student also agreed they would recommend the course to friends.
Students highlighted what they found most helpful in their learning: ``I loved the freedom given on the final project. I learned so much from just being able to explore." ``I liked the OpenCV course video that was assigned. It was fun going in-depth to a specific library and learning all you could do with it." Students also mentioned how, as someone familiar with programming already, the final projects and creating a demonstration helped students to self-learn topics needed for their project: ``For me, the most challenging topics were definitely learning the AI stuff." Some comments also reflected on the difficulty of balancing both CS and AI learning outcomes: ``I thought it was an awesome semester, but some of it seemed all over the place. Switching back and forth between topics was kind of hard to do and keep track of."

\begin{figure}[t!]
\centering
    \includegraphics[width=.405\textwidth]{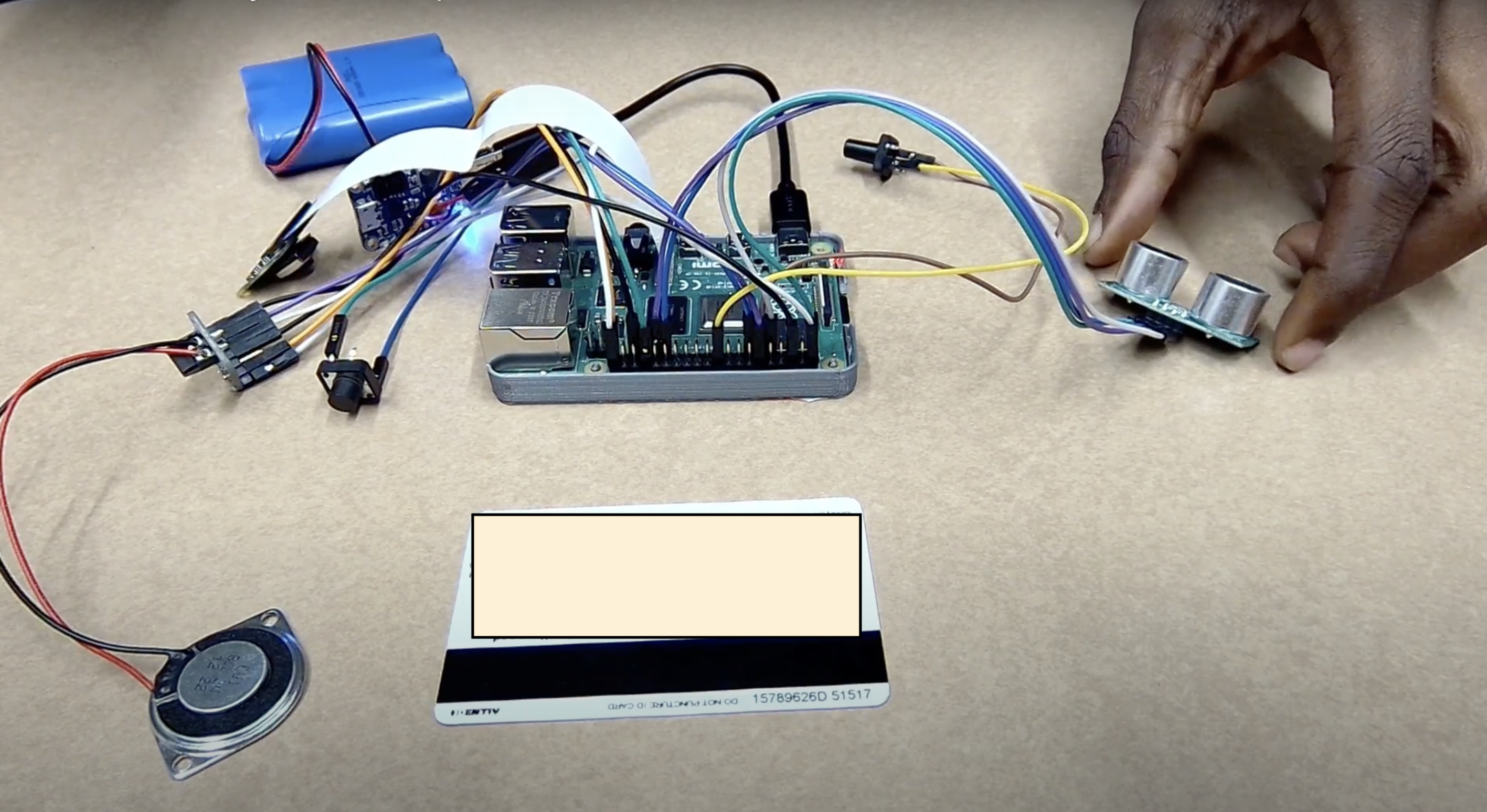}
    %\hspace{0.2pt}
    \includegraphics[width=.405\textwidth]{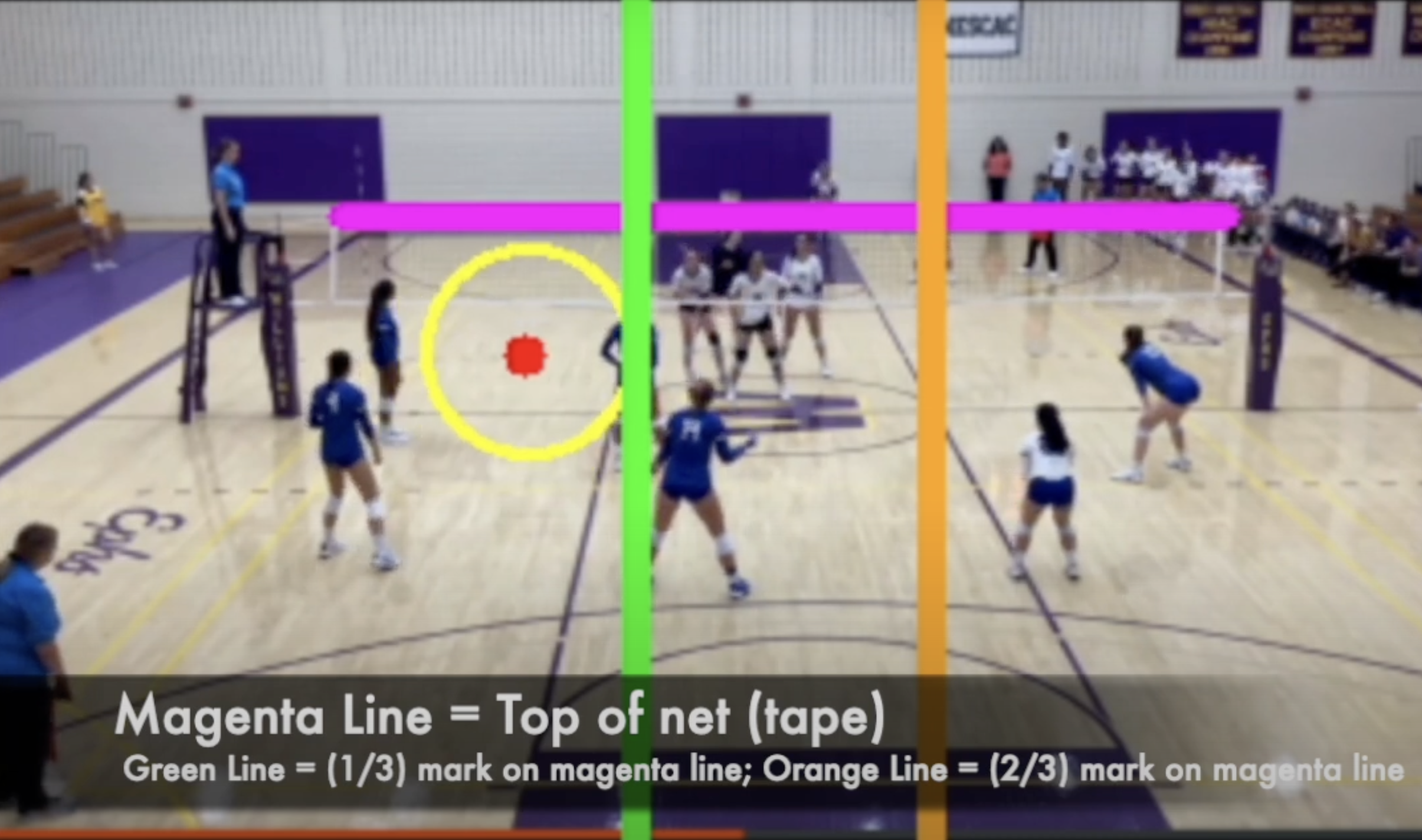}
    \caption{Images of two final project applications: Vision Assistant (left): a low cost reading assistive tool for visually impaired, and Passing Pro (right): a volleyball pass classification and analytics application.}
    %\Description[Radar plot showing aggregate student responses for the four repeated measures over the Fall semester for the CV.]{We observe an increase in AI Ethics Awareness, Perception of the environment, Sense of belonging and Self Efficacy, and decrease in Growth Mindset.}
\label{fig:project-screenshots}
%\vspace{-8pt}
\end{figure}

\subsection{Personal Project and Student Growth}

The most exciting aspect of the course for both students and the instructor was the final ``personal" project, where each student proposed, implemented, and demonstrated their work individually. The freedom granted in these projects was highly appreciated across the class. Student projects included a wide range of creative solutions: Vision Assistant (a low-cost device for the visually impaired to read books using Raspberry Pi, Optical Character Recognition, and text-to-speech models), Sport Scrutiny (a tool for analyzing and correcting sprint athletes’ running form), jaundice detection from skin photos using deep learning models, PassingPro (a volleyball pass classification and analytics program), hand2mouse (a tool that transforms hand motions into mouse controls), Virtual Zoo (an interactive web app to explore zoo animals), and an automatic license plate recognizer, among others.

Initially, we considered providing students with predetermined project topics, but allowing them to choose their own projects led to increased motivation and ownership, as students iterated and improved on something meaningful to them. The open-ended approach also resulted in a diverse array of project categories. Throughout the three deliverables, we scheduled time for students to form small groups, share their progress, and discuss challenges. This practice not only helped students understand their peers' projects but also fostered collaborative problem-solving. For many first-year students, this was a valuable opportunity to develop professional skills such as communication, cooperation, and research alongside the technical expertise needed for their future careers.

The intellectual challenge posed by the individual project set this course apart from other introductory CS courses for the students, as noted by one student: ``Every aspect of my final project was hard as I had to teach myself because I was ambitious with my project. It took me way longer than expected. Overall, I really enjoyed the course."

% Talk about the personal projects as part of learning growth
% some image

\section{Discussion and Instructor Reflections}

%The course outlined here was proposed with a set of learning outcomes designed for introductory students in a CS1 course who have some prior programming experience. The additional AI learning outcomes are designed to enable students to be familiar with some of the current methods used in computer vision and other AI sub-disciplines. 

From the first offering of this course in 2023, we identified several areas for improvement. Firstly, with the class meeting only twice a week, students spent 25-30 minutes each week working on their projects. While students appreciated the project topics, they felt these were more aligned with computer vision than with the lecture material. To address this, we plan to increase class meetings to three times a week (50 minutes each session), dedicating at least one session entirely to project work and in-depth discussions on AI reading assignments. Secondly, designing projects that integrate computer vision tools, weekly CS topics, and readings has proved to be challenging. In future iterations, we will incorporate participatory governance~\cite{10.1145/3626252.3630904} in project topic selection to better align projects with student interests and enhance their learning experience.

\textbf{Can this scale?} As this course was newly added to the catalog, we used two strategies to enroll students: a placement test for those with prior programming experience and transferring advanced students from other CS1-equivalent courses. Anticipating increased interest, which may strain teaching resources, we plan to hire teaching staff who have previously taken the course to provide additional support. With a large team of staff, the course could be offered during both regular and short semesters (4-14 weeks) at other institutions with larger classes. Currently offered for the second time with 20 students, we plan to implement these strategies to distribute the workload and support student projects effectively.

\textbf{How will students fare in the long term?} We recognize that this course's AI learning outcomes are not a substitute for more advanced AI courses. It is designed for students interested in AI while preparing to major in CS or other fields. One additional consideration is whether the learning outcomes of this course might affect students' CS1 knowledge and their success in subsequent CS-related courses compared to those on traditional CS pathways (CS1--CS2). In the future, we plan to compare performance in CS2 and other major courses for students from both CS1 versions (traditional and computer vision) to assess the impact on long-term learning and career success.

\textbf{Limitations.} This work was done in a small undergraduate-focused institution. This study reports results from the first offering of this course, and the placement test used to enroll advanced students in this introductory course is still under exploration. We also acknowledge that the small sample size ($n$ = 12) may limit the generalization of our findings to a larger class at a different institution.

% not a silver bullet
% how will they do in CS2 compared to other CS1 stydents, survey s may not reflect 

% Ca we do this for larger institutions
% what kind of knowledge they lack

%a lot fo students bring AI interest and blending that with CS can help with  attrition, retection, and recruitment
%since course was introduced for the first time, fewer students. Cna we scale
%What kind of knowledge they lack, tracking their progress in CS2

\section{Conclusion}

This paper documents the initial design of an introductory CS course aimed at fostering AI and computing knowledge. The course uses computer vision as an application context to enable learning outcomes for computational thinking and critical thinking. The students engaged with CS and AI through lectures, in-class programming activities, readings, and projects. A final individual project allowed the students to apply their learning to personally relevant problems. Students engaged in group activities such as debates and case study analysis based on the readings that encouraged critical discussions on AI's societal, ethical, and professional implications. Pre- and post-course survey data indicate improvement in students' sense of belonging, self-efficacy, and AI ethics awareness. Responses to reading activities show students' ability to inquire and analyze computing tools' impact on society and individuals. While preliminary evidence supports that the course fosters AI competence among students, we acknowledge that it cannot replace advanced AI courses but aims to spark further interest in AI and computing. We plan to evaluate the impact of this course on students' subsequent courses, majors, and career choices in the future.

% remaining survey questions
% remaining dimension figure
% reading list
\section{Acknowledgments}

\bigskip
\noindent We would like to thank Amanda Stent for the feedback on the design of this course. We also acknowledge helpful input from the teaching staff, students, and anonymous reviewers, which greatly improved this work. 

\bibliography{aaai25}

\end{document}